\documentclass[aps,prl,twocolumn,groupedaddress,amsmath,amssymb,showpacs,longbibliography]{revtex4-1}
\usepackage{graphicx,graphics,color}
\usepackage{amsmath}
\usepackage{amssymb}
\usepackage{amsfonts}
\usepackage{amsfonts}
\usepackage{epstopdf}
\usepackage{bm}
\usepackage{times,xspace}
\usepackage{xcolor}
\usepackage[utf8]{inputenc}
\usepackage{hyperref}

\def\be{\begin{equation}}
\def\ee{\end{equation}}
\def\bea{\begin{eqnarray}}
\def\eea{\end{eqnarray}}

\begin{document}

\title{Observation of non-Hermitian topology and its bulk-edge correspondence in an active mechanical metamaterial}

\author{Ananya Ghatak}
\affiliation{University of Amsterdam, Institute of Physics, Science Park 904, 1098 XH, Amsterdam, the Netherlands}
\author{Martin Brandenbourger}
\affiliation{University of Amsterdam, Institute of Physics, Science Park 904, 1098 XH, Amsterdam, the Netherlands}
\author{Jasper van Wezel}
\affiliation{University of Amsterdam, Institute of Physics, Science Park 904, 1098 XH, Amsterdam, the Netherlands}
\author{Corentin Coulais}
\affiliation{University of Amsterdam, Institute of Physics, Science Park 904, 1098 XH, Amsterdam, the Netherlands}

\date{\today}
\vspace{0.3cm}

\maketitle

\textbf{Topological edge modes are excitations that are localized at the materials' edges and yet are characterized by a topological invariant defined in the bulk. Such bulk-edge correspondence has enabled the creation of robust electronic, electromagnetic and mechanical transport properties across a wide range of systems, from cold atoms to metamaterials, active matter and geophysical flows. Recently, the advent of non-Hermitian topological systems---wherein energy is not conserved---has sparked considerable theoretical advances. In particular, novel topological phases that can only exist in non-Hermitian systems have been introduced. However, whether such phases can be experimentally observed, and what their properties are, have remained open questions. Here, we identify and observe a novel form of bulk-edge correspondence for 
a particular non-Hermitian topological phase. We find that a change in the bulk non-Hermitian topological invariant leads to a change of topological edge mode localisation together with peculiar purely non-Hermitian properties. Using a quantum-to-classical analogy, we create a mechanical metamaterial with non-reciprocal interactions, in which we observe experimentally the predicted bulk-edge correspondence, demonstrating its robustness. Our results open new avenues for the field of non-Hermitian topology and for manipulating waves in unprecedented fashions.}

\noindent
\emph{Keywords.} topological insulators, broken Hermiticity, mechanical metamaterials

\noindent
\emph{Significance statement.}
In recent years, the mathematical concept of topology has been used to predict and harness the propagation of waves such as light or sound in materials. However, these advances have so far been realized in idealised scenarios, where waves do not attenuate. In this research, we demonstrate that topological properties of a mechanical system can predict the localisation of waves in realistic settings where the energy can grow and/or decay. These findings may lead to novel strategies to manipulate waves in unprecedented fashions, for applications in vibration damping, energy harvesting and sensing technologies.
\section{Introduction}

\noindent
The inclusion of non-Hermitian features in topological insulators has recently seen an explosion of activity. Exciting developments include tunable wave guides that are robust to disorder \cite{Nash_PNAS2015, Mitchell_NatPhys2018, Khanikaev_NatComm2015}, structure-free systems \cite{Souslov_PRL2019,Delplace_Science2017}, and topological lasers and pumping \cite{Bandres_Science2018, Zilberberg_Nature2018, Kraus_PRL2012,Lohse_NatPhys2015,Pedro_PRL2019}. In these systems, active components are introduced to typically: (i) break time-reversal symmetry to create topological insulators with unidirectional edge modes \cite{Khanikaev_NatComm2015,Nash_PNAS2015,Souslov_PRL2019,Delplace_Science2017,Mitchell_NatPhys2018}; (ii) pump topologically protected edge modes, thus harnessing Hermitian topology in non-Hermitian settings \cite{Zeuner_PRL2015,Bandres_Science2018,Zilberberg_Nature2018,Kraus_PRL2012,Lohse_NatPhys2015}.
In parallel, extensive theoretical efforts have generalized the concept of a topological insulator to truly non-Hermitian phases that cannot be realised in Hermitian materials \cite{Gong_PRX2018, Shen_PRL2018,Ghatak_JPhysCondMat2019}.
However, such non-Hermitian topology and its bulk-edge correspondence remain a matter of intense debate. 
On the one hand it has been argued that the usual bulk-edge correspondence breaks down in non-Hermitian settings, while on the other hand new topological invariants specific to non-Hermitian systems have been proposed to capture particular properties of their edge modes~\cite{Yao_PRL2018,Kunst_PRL2018,Lee_PRB2019,Helbig_2019,Bergholtz_review2019,Xiao_NatPhys2020}.

\begin{figure}[b!]
\centering
 \includegraphics[trim=0cm 2.5cm 0cm 1cm, width=0.95\columnwidth]{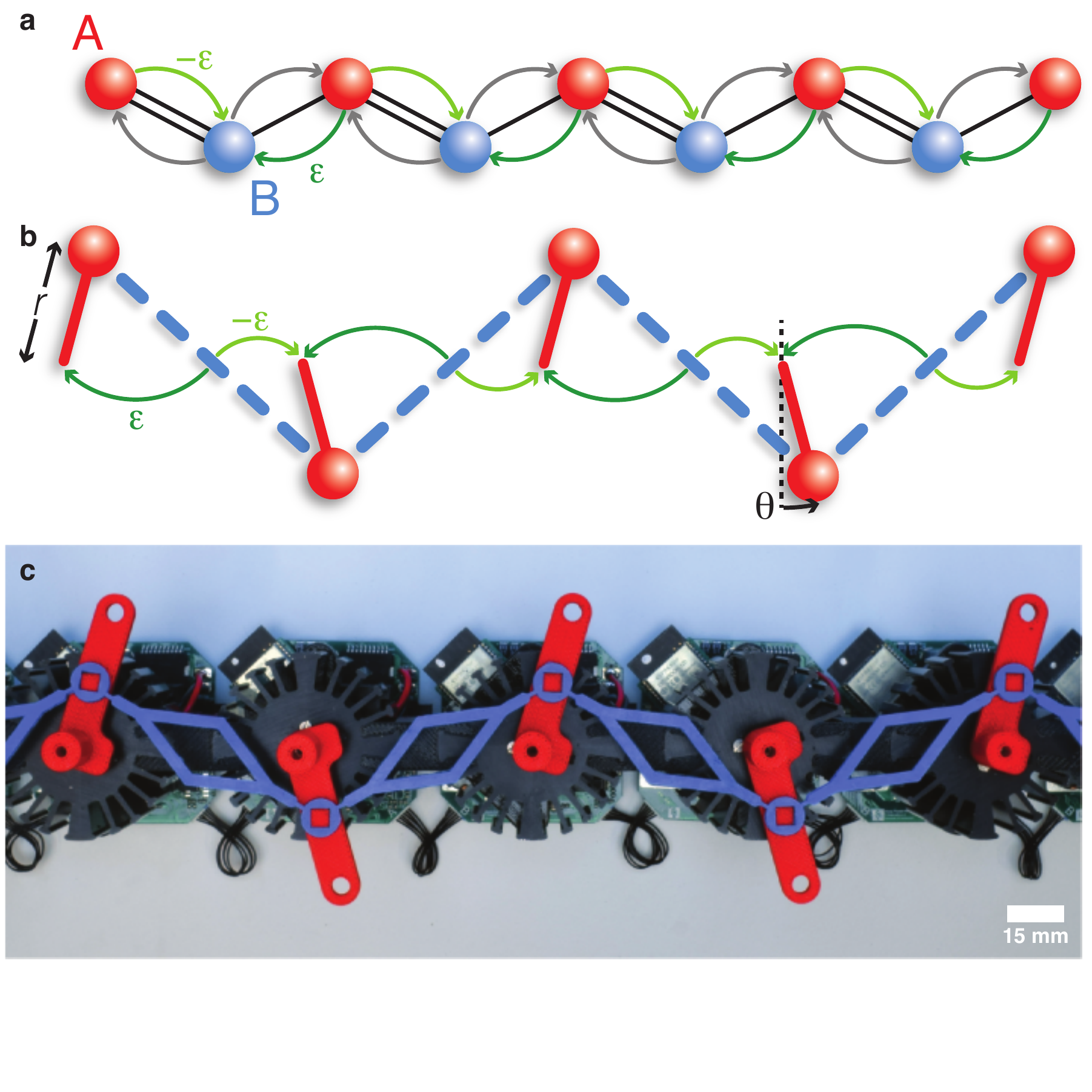}  \\
\caption{\textbf{Quantum-to-classical mapping of a chain with non-Hermitian topology.} (a): An SSH chain with two sublattices, A (in red) and B (in blue), augmented with non-reciprocal variations in the hopping amplitudes (indicated by $\pm \varepsilon$). (b): The non-reciprocal classical analogue of the augmented SSH chain, in which the classical masses (in red) correspond to the A sites in the quantum model, while the non-reciprocal springs (in blue) are analogous to the B sites. (c) Picture of the mechanical metamaterial realizing the non-reciprocal classical analogue of the augmented SSH model.}
\label{fig1}
\end{figure}

\begin{figure*}[t!]
\centering
 \includegraphics[width=2\columnwidth]{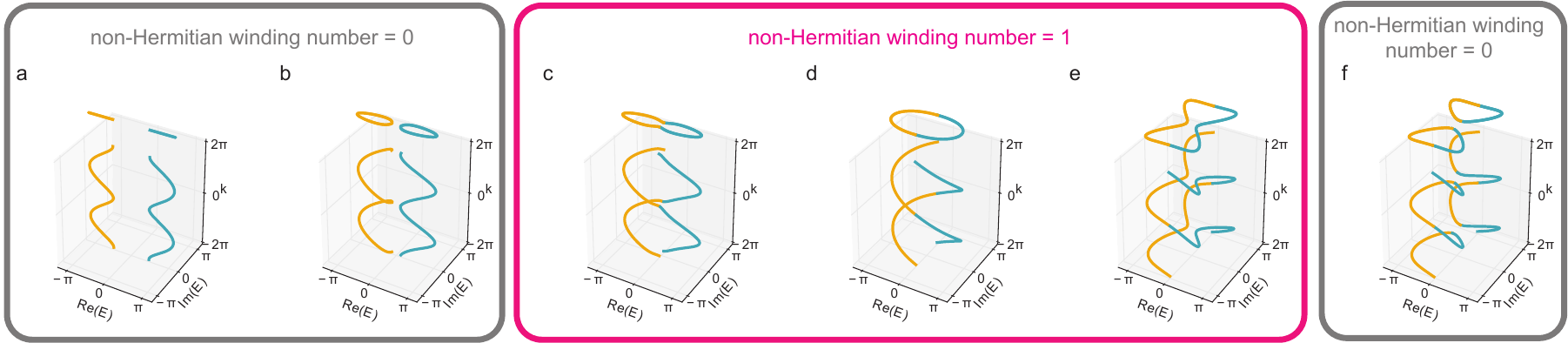}   \\
\caption{\textbf{Non-Hermitian topology.}  The real and imaginary parts of the energies (frequencies) of the two bands $E_-(k)$ ($\omega_-(k)$) (orange) and $E_+(k)$ ($\omega_+(k)$) (blue) as a function of wave number $k$, along with their projections onto the $k=0$ plane. The bands are shown for hopping parameters $a=2.5$ and $b=1$, and for six values of the non-reciprocal parameter $\varepsilon=0$, $0.4$, $0.45$, $0.9$, $2.3$, and $2.4$, grouped together into values corresponding from left to right to {non-winding} (a-b), {winding} (c-e), and again {non-winding} (f) non-Hermitian topology.}
\label{fig2}
\end{figure*}

Here, focussing on a non-Hermitian version of the Su-Schrieffer-Heeger (SSH) model \cite{MartinezAlvarez_PRB2018,Yao_PRL2018,Kunst_PRL2018,Lee_PRB2019} with an odd number of sites (Fig.~\ref{fig1}a), we find that a change in the bulk non-Hermitian topological invariant is accompanied by a localization change in the zero-energy edge modes. This finding suggests the existence of a
 bulk-edge correspondence for this new type of truly non-Hermitian topology. We further construct a mechanical analogue of the non-Hermitian quantum model (Fig.~\ref{fig1}b), and create a mechanical metamaterial (Fig.~\ref{fig1}c) in which we observe the predicted correspondence between the non-Hermitian topological invariant and the topological edge mode. In particular, we report that the edge-mode in the non-Hermitian topological phase has a peculiar nature, as it is localized on the rigid rather than the floppy side of the mechanical metamaterial.

\section{Non-Hermitian winding number}

\noindent
The one-dimensional model depicted schematically in Fig.~\ref{fig1}a, is described by the quantum mechanical Bloch Hamiltonian
\be
H(k) =\left(\begin{array}{ c c c c }
0  & Q(k) \\
R(k) & 0 \\ 
\end{array} \right),
\label{eq:SSH_Hamiltonian}
\ee
where $k$ is the wave vector. The coefficients $Q(k)= a_1+b_2 e^{-ik} $ and $R(k)=a_2+b_1 e^{ik}$ allow electrons to hop between neighbouring sites within the unit cell ($a_1$ and $a_2$), as well as between unit cells ($b_1$ and $b_2$). If the amplitudes for hopping to the left ($a_1$ and $b_1$) are different from the corresponding amplitudes for hopping to the right ($a_2$ and $b_2$), the Hamiltonian is non-Hermitian, with complex eigenvalues $E_\pm(k)=\pm\sqrt{Q(k) R(k)}$ that come in pairs related by reflection in the point $E=0$. Thus Eq.~(\ref{eq:SSH_Hamiltonian}) has a chiral symmetry, { ($\sigma_z^{-1}H(k) \sigma_z = -H(k)$)} and falls in symmetry class AIII \cite{Gong_PRX2018,Yoshida_arxiv2019}. 

A non-Hermitian Hamiltonian such as Eq.~(\ref{eq:SSH_Hamiltonian}) may host two different types of topological invariants, corresponding either to a winding of the phase of their eigenvectors as the wave vector $k$ is varied across the Brillouin zone \cite{Hasan_RMP2010} (see Eq.~(\ref{eq:Hwinding}) of the Materials and Methods), or to the complex energies winding around one another in the complex energy plane~\cite{Gong_PRX2018, Shen_PRL2018} (see Eq.~(\ref{eq:NHwinding}) of the Materials and Methods). The former type of topology exists both for Hermitian and non-Hermitian systems, while the latter is exclusive to non-Hermitian systems, has not been observed yet, and is the focus of the present work.
\begin{figure*}[t!]
\centering
 \includegraphics[trim=0cm 10cm 0cm 0cm,width=2\columnwidth]{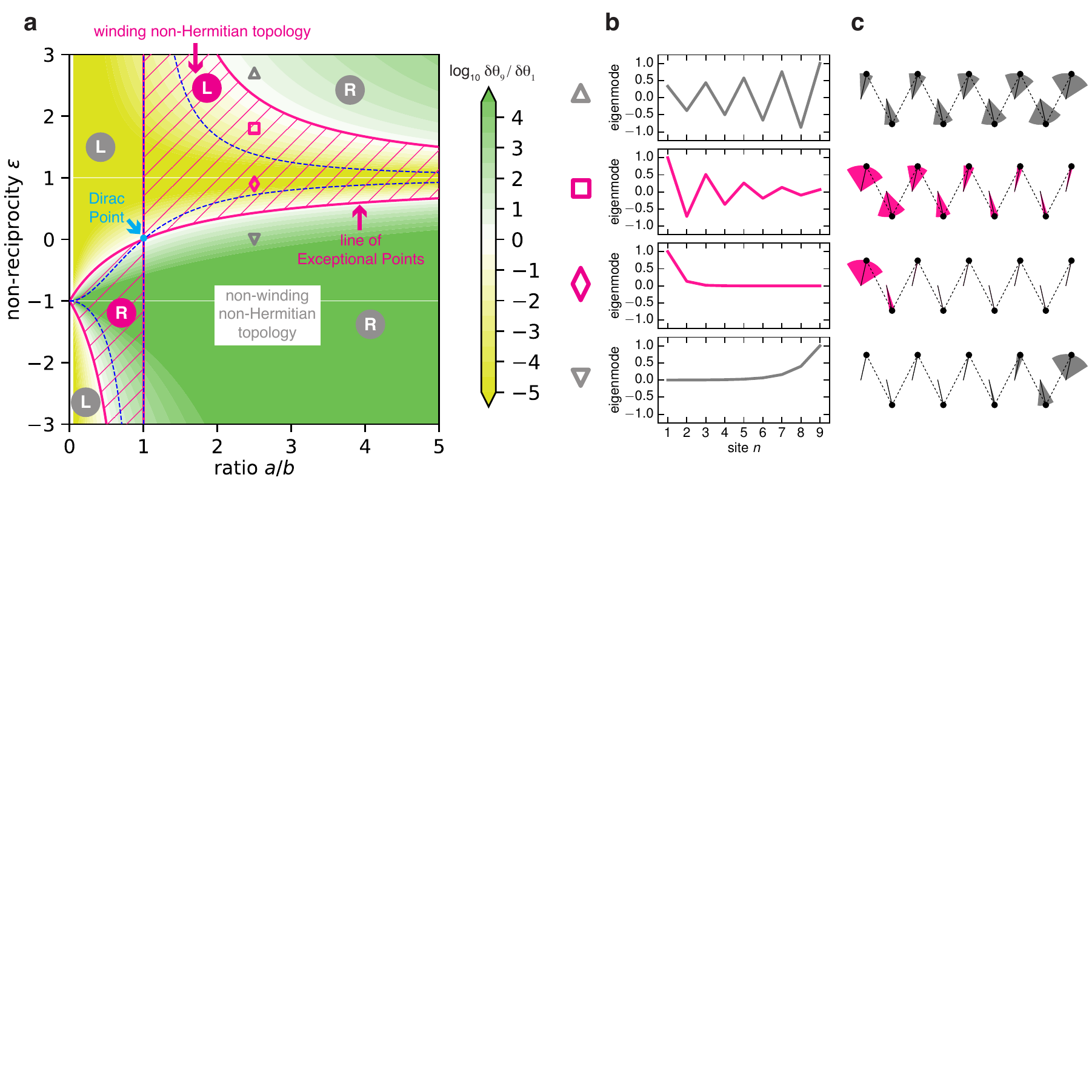} \\
\caption{\textbf{Phase diagram and bulk-edge correspondence.} (a) The phase diagram as a function of the parameters $a/b$ and $\varepsilon$. The hatched pink region corresponds to the topological phase with winding complex energies and $\nu=1$, while the other regions are topologically {non-winding} with two disconnected bands and $\nu=0$. The phase boundaries (thick solid pink lines) correspond to the bulk bands of the Bloch Hamiltonian touching the exceptional point $E=0$, at $\varepsilon=\varepsilon_c$. The thin dashed blue lines defined by $\varepsilon=\varepsilon_s$ (see Materials and Methods). 
The green contour plot represents the logarithm of the amplification factor of the zero edge mode for a chain of nine unit cells. The amplification factor is defined as $
|\psi_9^A/\psi_1^A|$ ($|\delta\theta_9/\delta\theta_1|$) in the quantum (mechanical) system (Materials and Methods), and indicates the side at which the zero mode is localised. The inverted triangle, diamond, square and triangle markers  correspond to parameters $a/b=2.5$ and $\varepsilon=0$, $0.9$, $1.8$, and $2.7$, respectively and indicate the parameters used for panels (b-c). The (L) and (R) labels indicate whether the topological edge mode is localized on the left or the right, respectively.
(b) Topological zero-energy modes of both the quantum model in Fig.~\ref{fig1}a with $9$ ($8$) $A$ ($B$) sites---evaluated only at $A$ sites, since the eigenmode at $B$ sites is zero---and of the classical model in Fig.~\ref{fig1}b with $9$ ($8$) rotors (springs). (c) Graphical representation of the zero-energy modes for the mechanical chain. The opening angle of the coloured wedges is proportional to the mode magnitude at each site.}
\label{fig3}
\end{figure*}

\section{Mapping between Non-Hermitian quantum and classical models} 

\noindent
The non-Hermitian topology contained in the model of Eq.~(\ref{eq:SSH_Hamiltonian}) stems from the non-reciprocity of its hopping parameters. This renders a direct implementation within a quantum material challenging, but recent advances on non-reciprocal mechanical metamaterials \cite{Nash_PNAS2015,Mitchell_NatPhys2018,Fleury_Science2014,Khanikaev_NatComm2015, Coulais_Nature2017,Wang_PRL2018,Brandenbourger_arxiv2019} suggest that such non-reciprocal interactions can be realized within a mechanical platform. In particular, inspired by the works of Kane and Lubensky \cite{Kane_NatPhys2014} and Brandenbourger et al.~\cite{Brandenbourger_arxiv2019}, we introduce the one-dimensional mechanical system (Fig.~\ref{fig1}b), which is described by the dynamical matrix:
\be
D(k)=(-a+b e^{-i k})(-a(1-\varepsilon )+b(1+\varepsilon )e^{i k}).
\label{eq:DynamicalMatrix_Fourier}
\ee
Here, $a=(p+2r\sin\theta)/\sqrt{p^2+4r^2\cos^2\theta}$ and $b=(p-2r\sin\theta)/\sqrt{p^2+4r^2\cos^2\theta}$ are geometrical parameters that depend on the length $r$, the initial angle $\theta$ of the red rotors and the lattice spacing $p$ (see Fig.~\ref{fig1}b and Materials and Methods). The parameter $\varepsilon$ modifies the stiffness of the blue springs in a non-reciprocal way, so that a strain in the spring causes a larger torque on the left rotor than on the right. { This non-reciprocal interaction is created locally for each robotic unit cell by an active-control loop: the motor of each unit cell applies a torque that depends on the strain of its neighbouring springs (see Materials and Methods).}

The equations of motion imposed by the dynamical matrix $D(k)$ on the displacements and their time derivatives may be combined into a Schr\"odinger-like equation, as proposed by Kane and Lubensky \cite{Kane_NatPhys2014, Lubensky_RepProgPhys2015,Huber_NatPhys2016,Susstrunk_PNAS2016}. The matrix taking the place of the Hamiltonian in this formulation has precisely the same form as Eq.~(\ref{eq:SSH_Hamiltonian}), with $Q(k)=-a+be^{-ik}$ and $R(k)=-a(1-\varepsilon)+b(1+\varepsilon )e^{ik}$ but the eigenvalues represent frequencies $\omega_\pm(k)$, rather than the energies (see Materials and Methods). This generalises the formal mapping between the dynamical matrix $D(k)$ and the Hamiltonian $H(k)$ introduced by Kane and Lubensky to a non-Hermitian setting.

\section{Non-Hermitian bulk-edge correspondence} 

\noindent
In the following, we restrict our attention to a particular model with parameter values $a_1=a, a_2=a(1-\varepsilon)$, $b_1=-b(1+\varepsilon)$ and $b_2=-b$. 
In the reciprocal and non-Hermitian limit of $\varepsilon=0$, the two bands of the model lie entirely on the real axis. As shown in Fig.~\ref{fig2}, increasing $\varepsilon$ leads to the bands developing imaginary components, and eventually touching at the exceptional point $E=0$ and $k=0$ before coalescing. The non-Hermitian winding invariant $\nu$ then {becomes} {one}. At even larger values of $\varepsilon$ another exceptional point is encountered at $k=\pi$ and the bands separate again into a {non-winding} phase.

We show the full phase diagram of this system in Fig. \ref{fig3}a, as a function of the hopping parameters $a/b$ and the non-reciprocal parameter $\varepsilon$. In the hatched pink region, the complex energies (frequencies) wind, and the non-Hermitian topological invariant has the value $\nu=1$. The other region has {non-winding} non-Hermitian topology with $\nu=0$, in accordance with the fact that the energies (frequencies) form disconnected bands. The phase boundaries correspond to the parameter values at which the bands coalesce at the exceptional point $E=0$~\cite{Shen_PRL2018,Gong_PRX2018,Ghatak_JPhysCondMat2019}, and are given by $\varepsilon_c=(a/b\pm 1)/(a/b\mp 1)$. The $\varepsilon=0$ axis represents the Hermitian limit, in which the non-Hermitian invariant is always zero and the two branches of exceptional points combine into a Dirac point at $a/b=1$. For the classical system, this axis corresponds precisely to the Kane-Lubensky model \cite{Kane_NatPhys2014}. The fact that the regions of non-Hermitian topology span large parts of parameter space suggests that they may be realized experimentally.

The hatched pink regions of the phase diagram Fig.~\ref{fig3}a are based on the behaviour of bulk topological invariants, calculated in a system with periodic boundary conditions. The non-Hermitian topology, however, is expected to be most visible experimentally in the emergence or suppression of edge modes localised at the edges of the chain. The edge modes can be found for the quantum (classical) model by solving Schr\"odinger's (Newton's) equation for zero-energy modes (see Materials and Methods). We focus in the following on a SSH chain with an odd number of sites (Fig.~\ref{fig1}a) and on the mechanical Kane-Lubensky chain (Fig.~\ref{fig1}b), for which the bulk-edge correspondences are strictly equivalent. Namely, we investigate a SSH (Kane-Lubensky) chain with $N$ A-sites (rotors) and $N-1$ B-sites (springs). {There is a vast literature on even-sized SSH chains and the choice of an odd-sized SSH chain can appear as less conventional. However, the topological nature of edge modes is the same for odd and even chains. Even in the Hermitian limit and for any given value of the topological invariant, it is the termination of the chain rather than the distinction between an odd and even number of atoms that determines the presence or absence of an edge mode. This can be clearly seen when considering for example the half-infinite chain, which is neither odd nor even and whose edge mode can be predicted to be present or absent based on knowledge of the invariant and the termination~\cite{history_topo}. Last but not least, choosing the odd chain is necessary to obtain a formal mapping with the mechanical system}.

In the Hermitian limit $\varepsilon=0$, both the quantum and the classical chain always have a single zero mode, which is localized to the right (left) edge for $a>b$ ($a<b$) (green contours in Fig.~\ref{fig3}a). In the non-Hermitian case $\varepsilon\neq 0$, the zero mode changes sides precisely at the critical lines $\varepsilon=\varepsilon_c$ of the bulk, periodic system (Fig.~\ref{fig3}a). In all cases, we find that the tails of the edge modes become oscillatory for $|\varepsilon|>1$ (Fig.~\ref{fig3}bc), as a consequence of imaginary contributions to their eigenvectors. Other choices of parameters will lead to a qualitatively similar correspondence between edge mode localisation and bulk winding (see Materials and Methods). Finally, we find that perturbations of the ideal model considered here, such as the inclusion of on-site potentials or mechanical bending interactions, progressively gap the system and suppress the zero-modes (see Materials and Methods).

{The coincidence between the change of the non-Hermitian winding number and the change of localisation of the zero modes demonstrates that these zero modes are topological and that a change of localisation corresponds to a topological transition.}  
{These topological zero modes have several peculiar properties, that are only possible because Hermiticity is broken:}
{(i) increasing the non-reciprocity at a fixed value of the ratio $a/b$ can be seen in Fig.~\ref{fig3}a to cause two consecutive changes in the edge mode location,  one of which goes against the direction of the non-reciprocal bias;} (ii) in the case of the quantum system, the shape of the phase diagram can not be explained by a simple argument involving the shifting of unit cells, as can be done in the Hermitian limit; (iii) in the mechanical system, as the topological edge mode in the winding region occurs where the mechanical degrees of freedom are constrained---this is a zero-energy mode, yet it involves stretching of the springs.

{The bulk-edge correspondence shown in Fig.~\ref{fig3}a differs from but is complementary to recent results on even non-Hermitian SSH chains, where the topological modes appear or disappear }
at the values $\varepsilon_s=((a/b)^2\pm 1)/((a/b)^2\mp 1)$ {at which the gap of the open chain is closed~\cite{Yao_PRL2018,Kunst_PRL2018}}. {That the gap closings of the open and closed chain do not coincide is a manifestation of the non-Hermitian skin-effect, which also causes all modes in the system except the topological zero mode to localise on one end of the chain
(see Materials and Methods)~\cite{MartinezAlvarez_PRB2018,Yao_PRL2018,Lee_PRB2019,McDonald_PRX2018,Brandenbourger_arxiv2019,Helbig_2019}.} 
{Recent results show that taking into account the non-Hermitian skin-effect allows the definition of a non-Bloch-topological invariant,  which  switches  value  at $\varepsilon_s$~\cite{Yao_PRL2018,Kunst_PRL2018} A  physically  compelling picture  thus  emerges  for  non-Hermitian  topological  phases: while the winding topological invariant predicts the edge at which the topological zero modes are localized,  the non-Bloch invariant predicts the existence of the topological modes and the location of the gap closing.} 

\begin{figure*}[t!]
\centering
\includegraphics[trim=0cm 0cm 0cm 0cm,width=1.8\columnwidth]{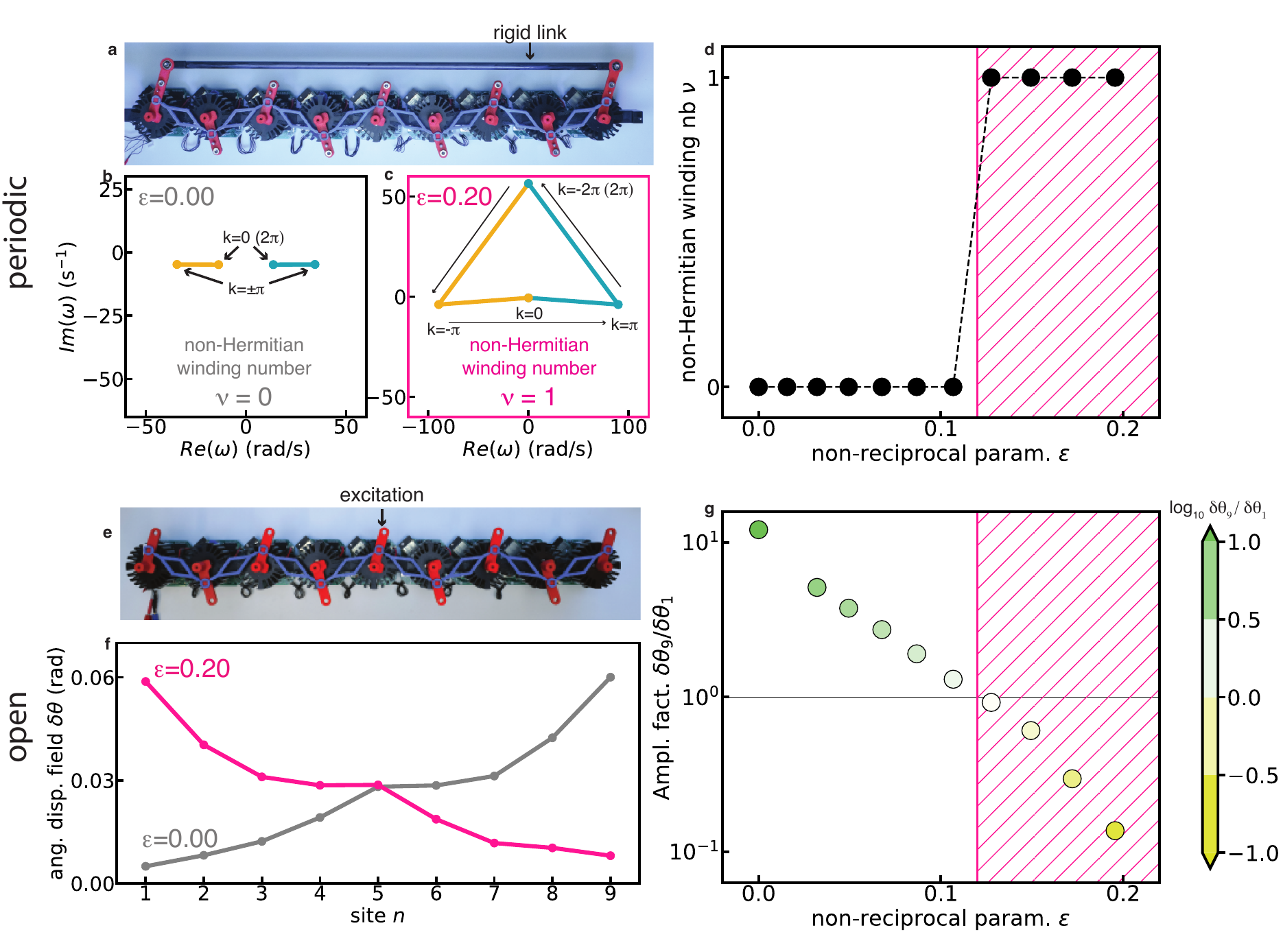} \\
\caption{\textbf{Experimental observation of the non-Hermitian bulk-edge correspondence.} (a) Picture of the periodic  metamaterial with nine unit cells, wherein the first and last rotors are rigidly connected. (b-c) Band diagrams showing the real part vs. the imaginary part of the eigenfrequencies for a non-reciprocal parameter $\varepsilon=0.00$ (b) and $\varepsilon=0.20$ (c). In (b) the two bands are disconnected ({non-winding} non-Hermitian topology), while in (c) they are connected ({winding}). (d) Corresponding measurement of the non-Hermitian winding number $\nu$ vs. non-reciprocal parameter $\varepsilon$. (e) Picture of the open metamaterial with nine unit cells. (f) Angular displacement field for different values of the non-reciprocal parameter $\varepsilon$, upon low-frequency excitation of the central unit for $\varepsilon=0.00$ (gray) and $\varepsilon=0.20$ (pink). (g) Amplification factor $\delta\theta_9/\delta\theta_1$ vs. non-reciprocal parameter $\varepsilon$. The hatched pink regions in (d) and (g) depict the non-hermitian winding phase for $\varepsilon>\varepsilon_c^\textrm{periodic}$ and $\varepsilon>\varepsilon_c^\textrm{open}$, respectively. The marker colormap quantifies the amplification factor, as in Fig. 3a. Details of the measurement protocols are in the  Materials and Methods. See also Supplementary Video 1.
}
\label{fig4}
\end{figure*}

\section{Non-Hermitian bulk-edge correspondence in {an active mechanical} metamaterial} 

\noindent
To demonstrate the non-Hermitian bulk-edge correspondence described above we provide an experimental realization. To this end, we build {an active mechanical} metamaterial (Figs. \ref{fig1}c and~\ref{fig4}a), which consists of nine robotic unit cells and in which a combination of geometry and active control is used to implement $D(k)$, as defined in Eq.~(\ref{eq:DynamicalMatrix_Fourier}). While the geometry allows us to obtain suitable values of $a=1.00$ and $b=0.73$, active control makes it possible to tune the non-reciprocal parameter $\varepsilon$ (see Materials and Methods). We selectively access properties of the periodic (bulk) or open (edged) system by either including or omitting a rigid connection between the first and last unit cells (Fig. \ref{fig4}a and Materials and Methods).
\begin{figure}[t!]
\centering
\includegraphics[trim=0cm 1cm 0cm 0cm,width=0.99\columnwidth]{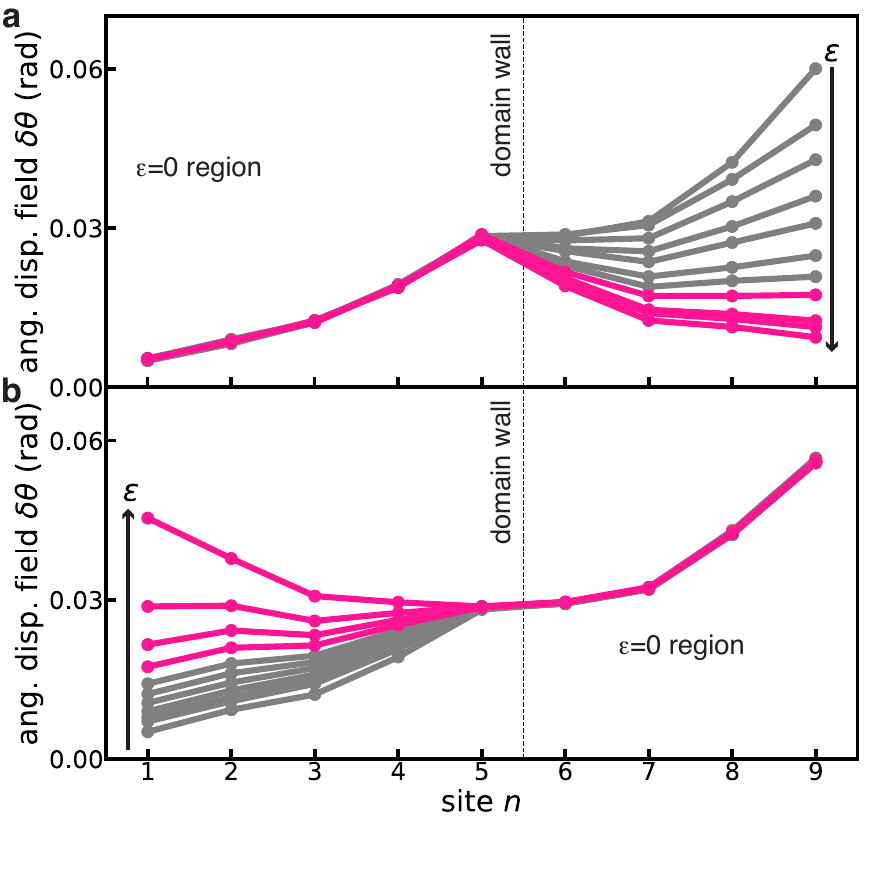}
\caption{\textbf{Domain Wall.} (a-b) Experimentally measured angular displacement field for different values of the non-reciprocal parameter $\varepsilon$, upon low-frequency excitation of the central unit. Data for $\varepsilon<0.12$ ($\varepsilon>0.12$) is shown in gray (pink). The metamaterial has a domain wall. The rightmost (a) or leftmost (b) part of the metamaterial is set to $\varepsilon=0$, while the leftmost (a) or rightmost (b) part of the metamaterial has the values of the non-reciprocal parameter $\varepsilon=[0.00,0.02,0.03,0.05,0.07,0.09,0.11,0.13,0.15,0.17,0.20]$.}
\label{fig4b}
\end{figure}

In this setup, we first perform relaxation experiments on the periodic metamaterial to quantify directly the bulk eigenfrequencies for the wave vectors $k=0,\pi$ and hence the bulk topological invariant $\nu$ (Fig. \ref{fig4}a-d). We find that the non-Hermitian topological invariant jumps from zero---where the eigenfrequencies are disconnected as in Fig. \ref{fig4}b---to one---where the eigenfrequencies wind as in Fig. \ref{fig4}c---for a non-reciprocal parameter $\varepsilon_c^\textrm{periodic}= 0.12$ (Fig. \ref{fig4}d). Second, we probe the signature of the zero modes of the open chain (Fig. \ref{fig4}e), by applying a low-frequency excitation at the central unit cell. We observe a right-to-left (left-to-right) decaying displacement field for small (large) values of the non-reciprocal parameter $\varepsilon$ (Fig. \ref{fig4}f and Supplementary Video). We find that the amplification factor $|\delta\theta_9/\delta\theta_1|$ crosses the value one at a critical value of $\varepsilon_c^\textrm{open}=0.12$ (Fig. \ref{fig4}g). Remarkably, the correspondence $\varepsilon_c^\textrm{periodic}=\varepsilon_c^\textrm{open}$ agrees precisely with the theoretically predicted non-Hermitian bulk-edge correspondence. It shows that the experimentally observable switching of edge state localization in the open chain coincides with the changing value of the non-Hermitian topological invariant in the (bulk) system with periodic boundary conditions. Moreover, it proves the robustness of both the bulk-boundary correspondence and the non-Hermitian topology to inherent deviations from the ideal model such as geometric and motor non-linearities, spring bending, time delays and noise in the micro-controllers, friction, and geometric irregularities. 

To show more clearly the connection between the topological transition and the behaviour of the edge modes, we also create a domain wall in the metamaterial, with the leftmost part remaining reciprocal ($\varepsilon=0$) and the non-reciprocal parameter $\varepsilon$ being tuned away from zero in the rightmost part (Fig. \ref{fig4b}c) and vice-versa (Fig. \ref{fig4b}d).  As expected, beyond the threshold value, the localization of the displacement field changes from the right edge to the domain boundary at the center (Fig. \ref{fig4b}c) or the displacement field localises at both edges away from the domain boundary (Fig. \ref{fig4b}d).

\section{Discussion and outlook} 

\noindent
To conclude, we discovered and experimentally observed a novel type of bulk-edge correspondence for the non-Hermitian topological phase of a mechanical metamaterial with non-reciprocal interactions. This particular form of non-Hermitian bulk-edge correspondence, connected to energy winding, exhibits marked differences with the recently proposed non-Hermitian bulk-edge correspondence connected to a bi-orthogonal expectation value~\cite{Yao_PRL2018,Kunst_PRL2018}. First, the correspondence based on energy winding reported here is unaffected by the non-Hermitian skin-effect: despite the complete reorganisation of the spectrum between a periodic and an open system, the energy winding of the periodic system predicts changes in the edge modes of the open system. 
Second, the energy winding and the bi-orthogonal condition both predict the emergence of zero modes. However, while the bi-orthogonal condition predicts the existence of edge modes, the energy winding additionally predicts the side of the chain at which the topological mode appears. 
These differences call for further investigation and generalization of the bulk-edge correspondence based on energy winding, beyond the particular system considered here.

Further, we envision the study of nonlinearity, robustness to disorder, different interactions, higher spatial dimensions and other strategies to achieve non-Hermiticity---to be exciting future research directions. We believe that our work provides conceptual and technological advances, opening up avenues for the topological design of tunable wave phenomena.

\setcounter{equation}{0}
\renewcommand{\theequation}{A\arabic{equation}}%




\section{Materials and Methods}

\noindent
\textbf{Hermitian and non-Hermitian topology of the non-reciprocal SSH model.}
The Hamiltonian of Eq.~(\ref{eq:SSH_Hamiltonian}) of the Main Text may acquire topological character either from the winding of the Berry connection determined by its eigenfunctions $|\psi^\pm(k)\rangle$, or from the direct winding of its eigenenergies $E^\pm(k)$ in the complex plane. While the former type of winding corresponds to conventional Hermitian topology~\cite{Lubensky_RepProgPhys2015}, the latter corresponds to a unique form of non-Hermitian topology, that can only exist when the eigenenergies are complex~\cite{Shen_PRL2018, Gong_PRX2018}.

The Hamiltonian in Eq.~(\ref{eq:SSH_Hamiltonian}) is pseudo-Hermitian with respect to a positive definite metric operator~\cite{Mostafazadeh_IJGMMP2010} and can be diagonalized to find its eigenenergies $E^\pm(k)=\pm\sqrt{R(k)Q(k)}$ as well as its left and right eigenmodes,
$
\langle\psi_L^+(k)|=\left(
1 , \sqrt{\frac{Q(k)}{R(k)}}\right)$, 
$\langle\psi_L^-(k)|=\left(-\sqrt{\frac{R(k)}{Q(k)}},
1 \right)$, 
$|\psi_R^+(k)\rangle=\left(1,\sqrt{\frac{R(k)}{Q(k)}}\right)^T$ 
and $|\psi_R^-(k)\rangle=\left(-\sqrt{\frac{Q(k)}{R(k)}},1 \right)^T$. 
These eigenmodes obey the bi-orthonormality condition and the Hamiltonian preserves a generalized unitarity condition~\cite{Mostafazadeh_IJGMMP2010}.

To compute the conventional Hermitian topological invariant in the non-Hermitian setting, we can define the Berry connection as~\cite{Ghatak_JPhysCondMat2019}
$
\mathcal{A}^\pm(k)=-i\langle \psi_L^\pm(k)\mid\partial_k \psi_R^\pm(k)\rangle, $
from which the topological invariant is then calculated to be
\be
\gamma^\pm =\frac{1}{2 \pi}\int_{0}^{2\pi}dk A^\pm(k). 
\label{eq:Hwinding}
\ee
Choosing different left and right eigenvectors in the definition of the Berry connection, 
or introducing a modified inner product, does not yield additional invariants~\cite
{Ghatak_JPhysCondMat2019}. The invariants $\gamma^\pm$ are zero (when $a/b >1$) or integer (when $a/b <1$) for the regions with {non-winding} non-Hermitian 
topology (non-hatched regions in Fig. \ref{fig3}a of the Main Text). In the {winding} non-Hermitian topological regions (pink hatched regions in Fig. \ref{fig3}a), the two bands coalesce into a $4\pi$-periodic structure. In that case the Berry connection winds an integer number of times after integration over the full $4\pi$ period, and we find $\frac{1}{2 \pi}\int_{0}^{4\pi}dk A^\pm(k)=1$ for the single, coalesced band.

The truly non-Hermitian topology can be defined in terms of the winding of eigenenergies around an exceptional point, as shown in Fig.~\ref{fig2} of the Main Text. The corresponding winding number can be calculated to be \be 
\nu=-\frac{1}{2 \pi}\int_{0}^{4\pi}dk \frac{\partial}{\partial k}(  \arg[E^+(k)-E^-(k)]).
\label{eq:NHwinding}
\ee
The value of this non-Hermitian topological invariant is shown in Fig.~\ref{fig3}a of the Main Text. Notice that the definition of Eq.~(\ref{eq:NHwinding}) differs by a factor of two from the convention used in some other works~\cite{Shen_PRL2018}.

\vspace{1eM}
\noindent
\textbf{Edge modes of the non-reciprocal SSH model with an odd number of sites.}
In Eq.~(\ref{eq:SSH_Hamiltonian}) of the Main Text, a non-reciprocal version of the SSH model is defined in reciprocal space. Here, we use the corresponding real space formulation to identify the edge modes of a finite non-reciprocal SSH chain with open boundary conditions. Specifically, we consider $N$ sites of type A, and $N-1$ sites of type B, which is strictly analogous to the mechanical Kane-Lubensky chain~\cite{Kane_NatPhys2014}. The Hamiltonian is given in real space by:
\begin{equation}
{\bf H}=\left(
\begin{array}{ccccccc}
 0   & a_1&   0 &    & \dots&   & 0\\
 a_2 &   0& b_1 &    & &   & \\
 0   & b_2&   0 & a_1 & \ddots&   & \vdots\\
    &   & a_2 &   0 & \ddots& & \\
 \vdots   &   &    \ddots& \ddots & \ddots&  a_1 & 0\\
    &  &    &  & a_2&  0 & b_1\\
 0   &   &   \dots &    &0 &b_2 & 0\\
\end{array} 
\right),
\end{equation}
where $a_1$, $a_2$, $b_1$, and $b_2$ are hopping coefficients, which we assume to be real. In the Hermitian case, with $a_1=a_2$ and $b_1=b_2$, the Hamiltonian has a unique zero mode satisfying the equation ${\bf H} |\psi\rangle={\bf 0}$, with the eigenmode written as $|\psi\rangle=(\psi^A_1,\psi^B_1,\dots,\psi^A_{N-1},\psi^B_{N-1},\psi^A_{N})^T$. However, in the generic non-Hermitian case considered here, with $a_1\neq a_2$ and $b_1\neq b_2$, the equations ${\bf H} |\psi_R\rangle={\bf 0}$ and $\langle\psi_L| {\bf H}={\bf 0}$ respectively for the right and left eigenvectors may yield distinct zero-energy modes. We solve these two equations and find $(\psi^A_n)_R/(\psi^A_1)_R =(-a_2/b_1)^{n-1}$, $(\psi^B_n)_R=0$ for the right eigenmode, and $(\psi^A_n)_L /(\psi^A_1)_L =(-a_1/b_2)^{n-1}$, $(\psi^B_n)_L=0$ for the left eigenmode. In Fig.~\ref{fig3}b of the Main Text, we only plot the right eigenmodes for $a_1=a$, $a_2=a(1-\varepsilon)$, $b_1=-b(1+\varepsilon)$ and $b_2=-b$.

\begin{figure}[t!]
\centering
 \includegraphics[ width=0.9\columnwidth]{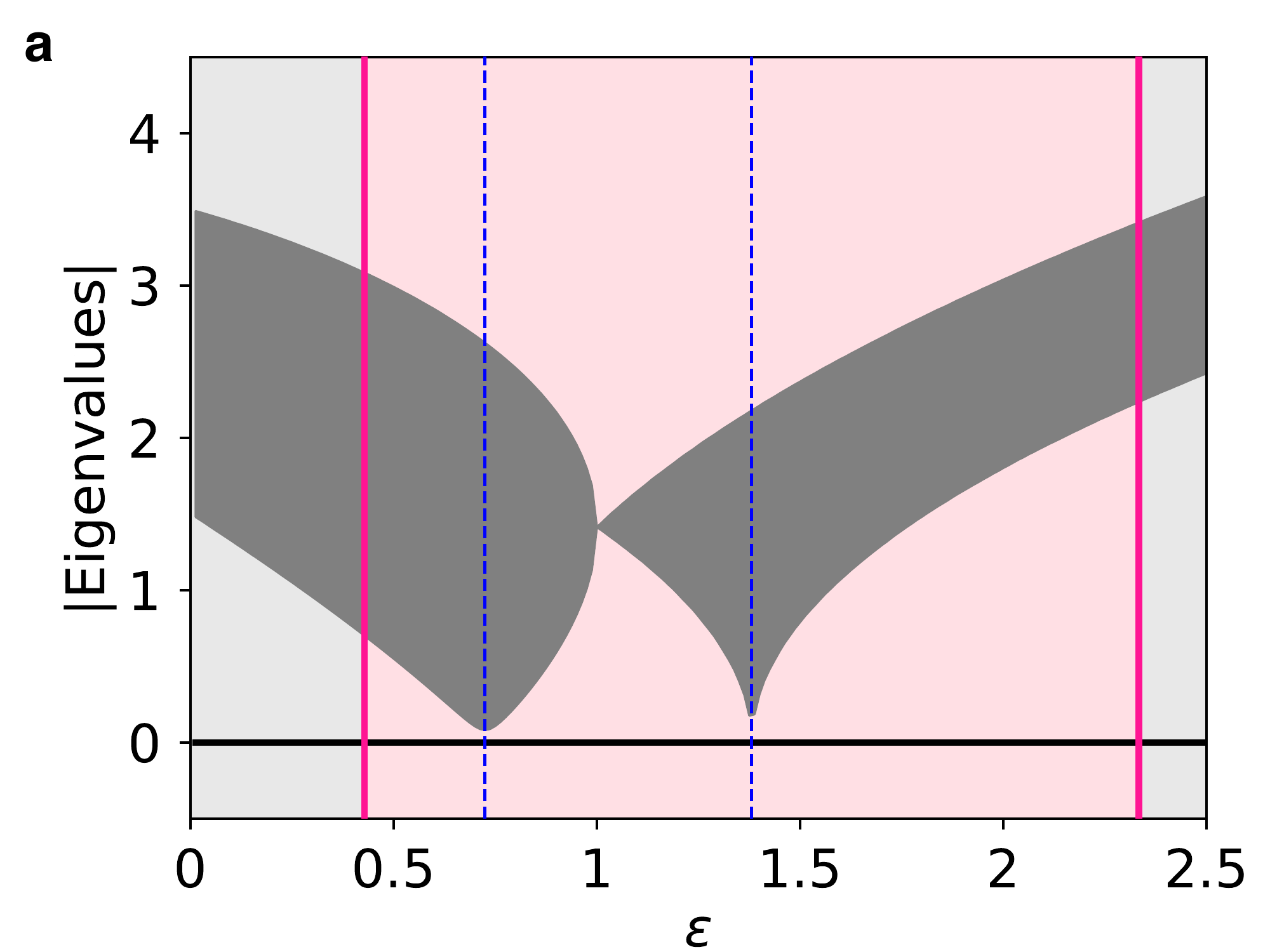}  \\
  \includegraphics[ width=0.9\columnwidth]{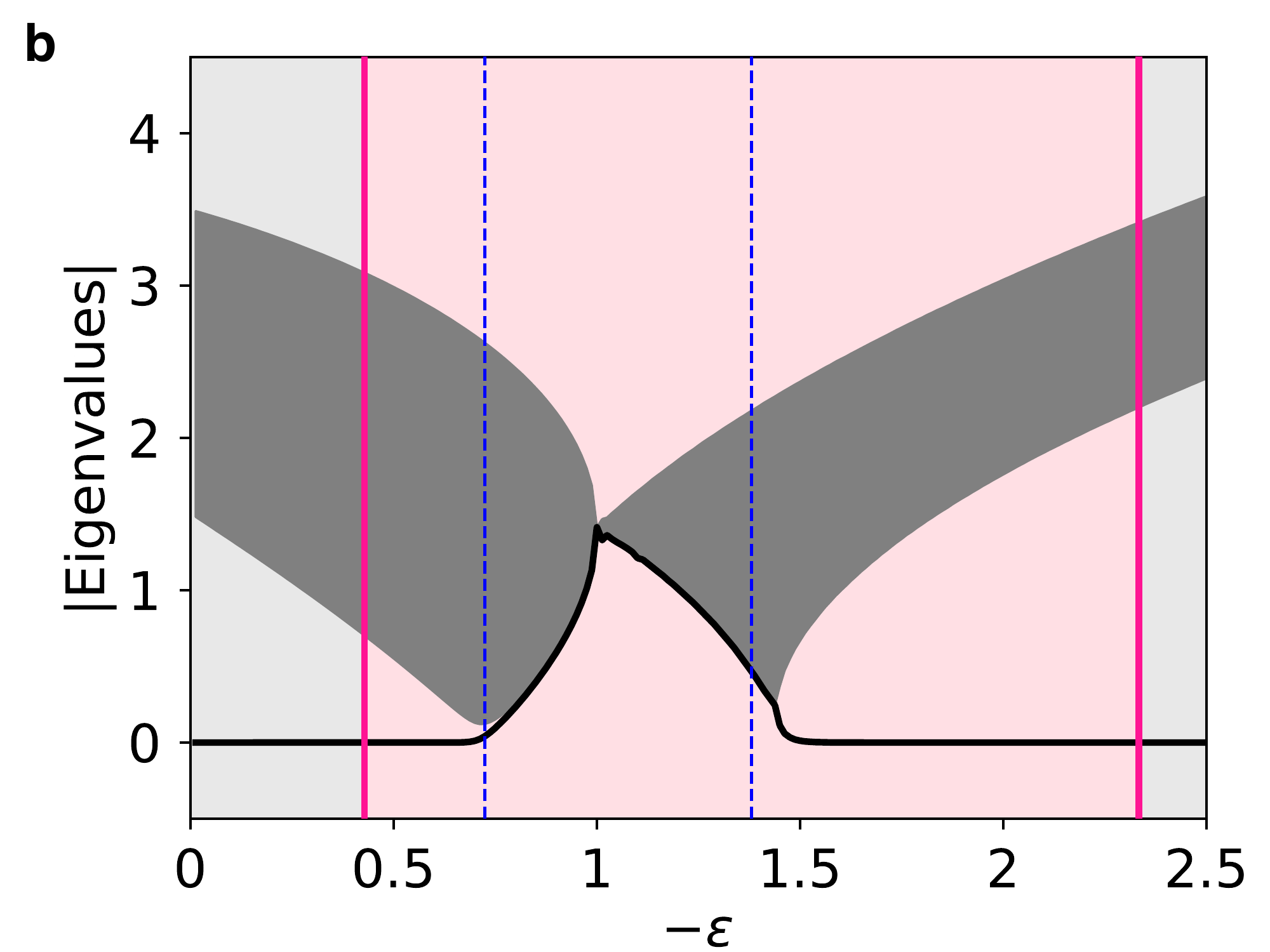}  \\
\caption{\textbf{Spectra with open boundary conditions.} (a) {Eigenvalues of an odd finite non-Hermitian SSH chain with $50$ A sites and $49$ B sites with open boundary conditions, plotted as a function of the non-reciprocal parameter $\varepsilon$.} Here we used parameters values $a=2.5$ and $b=1$, as in Figs.~\ref{fig2} and \ref{fig3} of the main text. The black line corresponds to the lowest energy mode and the gray band to the region enclosing the rest of the spectrum. The pink region is delimited by thick solid pink lines and indicates the values of $\varepsilon$ for which the bulk energies (computed with periodic boundary conditions) wind. The zero-energy mode of the open chain changes its localization as $\varepsilon$ is tuned into or out of this region. The dashed blue lines show the values of $\varepsilon$ for which the spectrum of the open chain become gapless, and correspond to a jump of the non-Bloch invariant defined in \cite{Yao_PRL2018}. {(b) Eigenvalues of an even finite non-Hermitian SSH chain with $50$ A sites and $50$ B sites. Here, we used $a=1$ and $b=2.5$, to ensure that an edge state exists when the non-reciprocal parameter $\varepsilon$ crosses the critical values indicated by thick pink lines. The indicated regions and lines have the same meaning as in the top panel, but in this case, the lowest energy mode is always doubly degenerate, with one mode changing localization at the thick pink lines. That the gap closing in the open chain does not seem to coincide with the blue dashed line is due to the finite size of the chain used.}
}
\label{figA1}
\end{figure}

\vspace{1eM}
\noindent
\textbf{Non-Hermitian skin effect.} In the non-Hermitian case, boundary conditions have a significant effect on the shape of the entire spectrum, namely open boundaries shift modes at all energies (frequencies) towards one side of the chain, in what is known as the non-Hermitian skin effect~\cite{Yao_PRL2018,Lee_PRB2019,McDonald_PRX2018,Borgnia_arxiv2019}. This effect is not related to topology, and was recently observed in both a non-reciprocal mechanical metamaterial~\cite{Brandenbourger_arxiv2019} and a non-reciprocal electronic circuit~\cite{Helbig_2019}. 
Theoretically, it has been shown that in a non-Hermitian SSH chain, the closing of the bandgap appears at parameter values that are different for open and periodic boundary conditions, in an apparent breakdown of the bulk-edge correspondence~\cite{Kunst_PRL2018,Yao_PRL2018,Lee_PRB2019,Borgnia_arxiv2019} {(indicated by the grey region in Fig.~\ref{figA1}b)}. 
We find consistent results (indicated by the grey region in Fig.~\ref{figA1}a), but we report in addition a clear correspondence between the topology of the bulk spectrum computed with closed boundary-conditions, and the zero-energy edge modes obtained with open boundary conditions: (i) a zero-energy edge mode always exists, as calculated analytically in the section above and confirmed numerically (Fig.~\ref{figA1}a); (ii) this edge mode in the chain with open boundaries changes its localization at the exceptional points of the periodic---bulk--- model (See Fig.~\ref{fig3}ab of the Main Text); (iii) surprisingly, this edge mode is unaffected by the gap closing $\varepsilon=(a^2/b^2\pm 1)/(a^2/b^2\mp 1)$ of the open chain~\cite{Yao_PRL2018} (dashed lines in Fig.~\ref{fig3}a of the Main Text and Fig.~\ref{figA1}). {These results are consistent with and complementary to recent results in the case of the even non-Hermitian SSH chain~\cite{Kunst_PRL2018,Yao_PRL2018}}

\begin{figure}[t!]
\centering
 \includegraphics[ width=0.95\columnwidth]{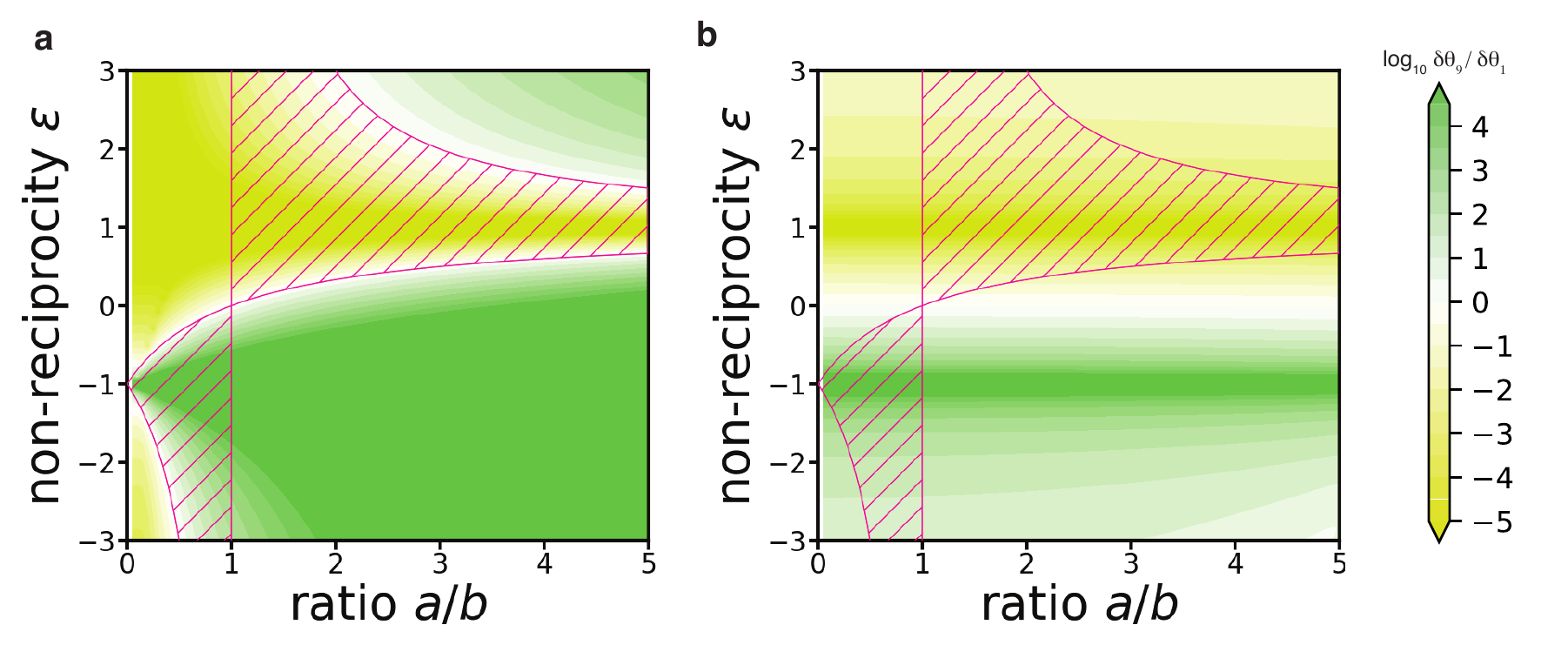}  \\
\caption{\textbf{Signature of the non-Hermitian skin effect.} Amplification factor for an excitation at a dimensionless radial frequency $\omega=0.01$ (a) and $\omega=10$ (b) for the metamaterial. The hatched pink region depicts the non-Hermitian topological phase. Notice that the localisation of the low-frequency topological response is opposite to that of the high-frequency skin effect in extensive regions of parameter space.}
\label{figA1b}
\end{figure}

Notice that in our system, a signature of the non-Hermitian skin effect can be seen in the response to a local excitation. While the localization of the response at low frequency (Fig.~\ref{figA1b}a) is essentially the same as that of the zero-mode (Fig. 3 of the main text), changing localization edge at the topological phase boundaries, the localization at large frequencies (Fig.~\ref{figA1b}b) solely depends on the non-reciprocal parameter, which is a direct signature of the non-Hermitian skin effect. For extensive portions of parameter space, the localization of the topological zero mode dominating the low-frequency response is opposite to that induced by the non-Hermitian skin effect at high frequencies.

\vspace{1eM}
\noindent
\textbf{Non-reciprocal Kane-Lubensky chain.}
The classical analogue of the non-Hermitian SSH chain is a non-reciprocal version of the Kane-Lubensky chain~\cite{Kane_NatPhys2014}, as shown in Fig.~\ref{fig1}b of the Main Text. In this system, $N$ rotors of length $r$, with an initial tilt angle $\theta$ and a staggered orientation, are connected by $N-1$ springs between subsequent rotors, with lattice spacing $p$. 
To construct the equation of motion for such a system, we first write the relation between the angular displacements of the rotors, $|\delta\theta\rangle=(\delta\theta_1,\dots,\delta\theta_N)^T$, and the length change of the springs (positive for stretching, negative for compression), $|\delta\ell\rangle=(\delta\ell_1,\dots,\delta\ell_{N-1})^T$. This is given by 
$|\delta\ell\rangle={\bf R} |\delta\theta\rangle$, with the compatibility matrix~\cite{Lubensky_RepProgPhys2015}:
\begin{equation}
{\bf R} = r \cos \theta\left(\begin{array}{ c c c c c}
-a  &  b  & 0 & \dots & 0 \\
0 & -a  &  b &  & \vdots \\
\vdots  &  & \ddots  & \ddots & 0 \\
 0 & \dots & 0 & -a &  b
\end{array} \right). \label{eq:comp}
\end{equation}
The coefficients $a= (p- 2r\sin\theta)/\sqrt{p^2+4r^2\cos^2\theta}$ and $b=(p+ 2r\sin\theta)/\sqrt{p^2+4r^2\cos^2\theta}$ are geometric parameters
~\cite{Kane_NatPhys2014}. We can similarly write the relation between the torque on each rotor, $|\tau \rangle=(\tau_1,\dots,\tau_N)^T$, and the tension in the springs, $|t \rangle=(t_1,\dots,t_{N-1})^T$, in the form $|\tau \rangle={\bf Q} |t \rangle$, where ${\bf Q}$ is the equilibrium matrix. In the Hermitian case ${\bf R}$ and ${\bf Q}$ are transposes of each other~\cite{Lubensky_RepProgPhys2015}. The compatibility and equilibrium matrices can be multiplied to compute the so-called dynamical matrix $\bm{D}^{\textrm{Hermitian}}=\frac{k}{J}\bm{Q}\bm{R}$, where $k$ is the spring constant of the elastic link between subsequent rotors and $J$ is the rotational moment of inertia of the rotors. 

In the {active mechanical} metamaterial, each unit cell $n$ has a local control system that can apply a local torque $\tau^{\textrm{control loop}}_n$ that depends on the angular displacement of its rotor $\delta\theta_n$ and that of its neighbours $\delta\theta_{n-1}$ and $\delta\theta_{n+1}$. We choose to apply the following torque
$\tau^{\textrm{control loop}}_n= \varepsilon k r^2\cos^2\theta (b (b\delta\theta_n-a\delta\theta_{n-1}) -  a (a\delta\theta_n-b\delta\theta_{n+1}))$. Since the added torques are proportional to the angular displacements, the system can still be described by an effective dynamical matrix $\bm{D}=\bm{D}^{\textrm{Hermitian}}+\bm{D}^{\textrm{control loop}}$ of the form
\begin{widetext}
\be
{\bf D}= \frac{k r^2\cos^2\theta}{J}\left(\begin{array}{ c c c c c c c }
a^2 (1-\varepsilon ) & -ab(1+\varepsilon )  & 0 & 0 & - & - & 0 \\
-ab(1-\varepsilon ) & a^2(1-\varepsilon )+b^2(1+\varepsilon )& -ab(1+\varepsilon ) & 0 & - & - & 0  \\
0 & -ab(1-\varepsilon ) &a^2(1-\varepsilon )+b^2(1+\varepsilon ) & -ab(1+\varepsilon ) & - & - & 0  \\
 & & \ddots & \ddots & \ddots & & \\
0 & 0  & - & - & 0 & -ab(1-\varepsilon ) & b^2(1+\varepsilon ) 
\end{array} \right)
\label{D},
\ee
\end{widetext}
which is non-symmetric for $\varepsilon\neq 0$.
In the Main Text, we assume that the ratio $k r^2\cos^2\theta/J=1$ without loss of generality. 
 
We compute the right and left zero modes by solving ${\bf D}|\delta\theta_R \rangle={\bf 0}$ and $\langle\delta\theta_L | {\bf D}={\bf 0}$, and we find $(\delta\theta_n)_R/(\delta\theta_1)_R=(a(1-\varepsilon)/b(1+\varepsilon))^{n-1}$ and $(\delta\theta_n)_L/(\delta\theta_1)_L=(a/b)^{n-1}$, respectively. We show and discuss only the right zero modes in Fig.~\ref{fig3}bc and in the Main Text, because  the right zero modes dominate the observed angular displacement profile~\cite{Schomerus}.

With periodic boundary conditions, the Fourier transform of Eq.~(\ref{D}) becomes
$D(k)=a^2(1-\varepsilon )+b^2(1+\varepsilon) -ab(1+\varepsilon )e^{ik}-ab(1-\varepsilon )e^{-ik}$, which in its factored form $D(k)=Q(k) R(k)$, with $Q(k)=(-a+b e^{-ik})$, $R(k)=(-a(1-\varepsilon)+b(1+\varepsilon) e^{ik})$, coincides with Eq.~(\ref{eq:DynamicalMatrix_Fourier}) of the Main Text. This factored mathematical expression allows us to construct the mapping between quantum and classical systems, following References~\cite{Kane_NatPhys2014,Lubensky_RepProgPhys2015,Huber_NatPhys2016,Susstrunk_PNAS2016}, where the quantum Hamiltonian is written as in Eq.~(\ref{eq:SSH_Hamiltonian}) of the Main Text.
Notice that the physical meaning of the Fourier equilibrium and compatibility matrices $Q(k)$ and $R(k)$ is ill-defined in the non-Hermitian case.

\begin{figure}[b!]
\centering
 \includegraphics[trim=0cm 2cm 0cm 0cm, width=0.99\columnwidth]{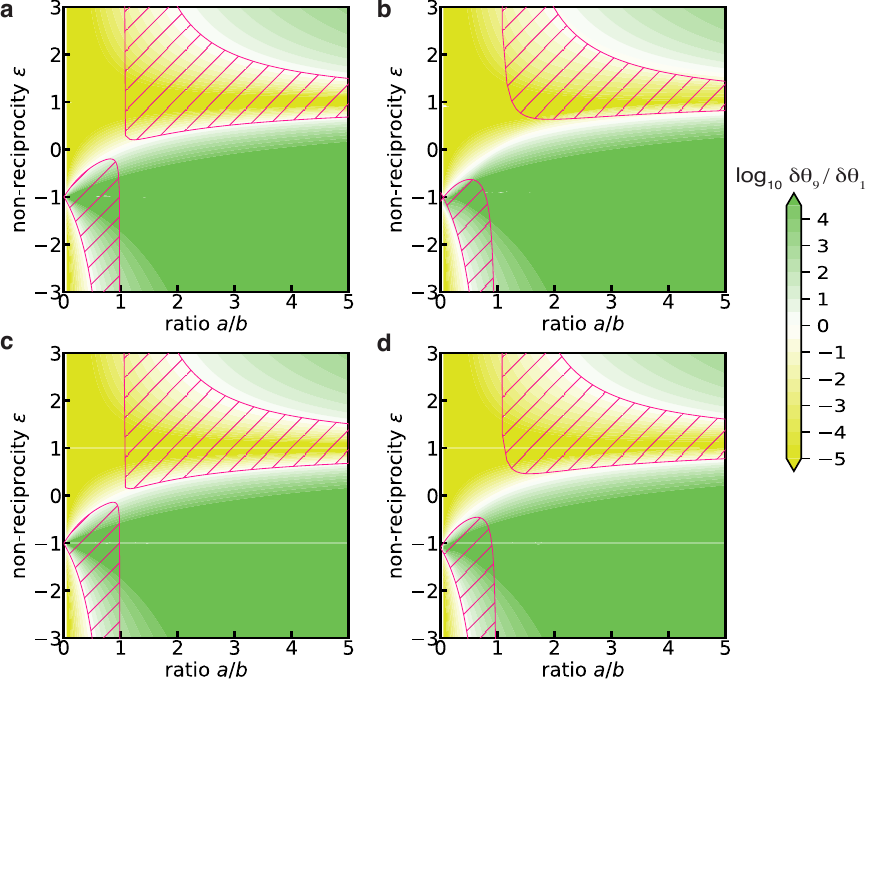}  \\
 \caption{\textbf{Role of perturbations}. The evolution of the phase diagram as a function of bending (a-b) and on-site potential (c-d). The phase diagram is shown for two different values of the bending stiffness parameter $\rho=0.01$ (a) and $\rho=0.1$ and two different values of the on-site potential $g=0.01$ (c) and $g=0.1$ (d). The hatched pink region indicates the phase with {winding} non-Hermitian topology given by Eq.~(\ref{eq:onsite_bending}), in which the model with periodic boundary conditions has winding eigenvalues. The color scale shows the amplification factor in the model with open boundary conditions.}
\label{figA2}
\end{figure}

\vspace{1eM}
\noindent
\textbf{Role of perturbations}
The computational model introduced in Eq.~(\ref{D}) and discussed in the Main Text is necessarily an idealisation. In the actual mechanical, integrated system in the experimental setup there are unavoidable small effects of bending in each of the elastomeric bands, on top of other essential effects from frictional forces, geometrical and electromechanical non-linearities in the chain, time delays and noise from micro-controllers and geometric irregularities. That we nevertheless see the theoretically predicted bulk-edge correspondence in our experimental results is thus witness to the robustness of the non-Hermitian topology described by the numerical predictions.

To test the limits of the topological robustness, we explicitly probe the role of two types of perturbations in the numerical model:  (i) bending interactions inherently present in the elastomeric bands connecting the nearest neighbour rotors; (ii) an on-site potential.
With these, the Fourier-transformed dynamical matrix becomes $D(k)=a^2(1-\varepsilon +\rho+g)+b^2(1+\varepsilon+\rho+g) -ab(1+\varepsilon -\rho)e^{ik}-ab(1-\varepsilon -\rho)e^{-ik}$], where $\rho$ is the relative bending stiffness of each rubber band and $g$ is the on-site potential. For the system with periodic boundary conditions, described by $D(k)$, the phase boundaries between {non-winding} non-Hermitian topology and {winding} non-Hermitian topology become
\begin{align}
\varepsilon_c=\frac{a/b\pm 1}{a/b\mp 1}+\rho\frac{a/b\mp 1}{a/b\pm 1}+g\frac{(a/b)^2+ 1}{(a/b)^2- 1}.\label{eq:onsite_bending}
\end{align}
Fig. \ref{figA2}ab shows that the presence of bending gaps the spectrum of the model with periodic boundary conditions around $\varepsilon=0$ for all values of the ratio $a/b$. At higher values of non-reciprocal parameter $\varepsilon$, however, the transition into a phase with non-Hermitian topology survives. Likewise, for non-zero on-site potential the gapping of the spectrum in the periodic chain causes a breakdown of the bulk-boundary correspondence around the point $a/b=1$ and $\varepsilon=0$, while for higher values of $\varepsilon$ it survives (Fig. \ref{figA2}cd). Both of these observations are a testament to the robustness of the non-Hermitian topology and its bulk-boundary correspondence, which remain intact even in the presence of perturbations, for sufficiently large values of the non-reciprocity. These considerations have been taken into account in the design of the experiments described below, wherein a specific shape of the rubber band is chosen to minimize the bending.

\vspace{1eM}
\noindent
\textbf{Experimental platform}
To perform the experiments, we followed Brandenbourger et al.~\cite{Brandenbourger_arxiv2019}, and created a one-dimensional, non-reciprocal {active mechanical} metamaterial (Fig.~\ref{fig1}c) consisting of nine unit cells, each of which has a single rotational degree of freedom $\theta_n$, where $n$ is the unit cell index. 
The unit cells are mechanically connected in a specific geometry initially proposed in the paper of Kane and Lubensky~\cite{Kane_NatPhys2014} (Fig.~\ref{fig1}b). Each unit cell consists of a rigid rotor of length $36$\,mm, and initial angle $(-1)^n(\theta-\pi/2)$ with respect to the horizontal axis, where $\theta=\pi/12$. Each rotor is connected to its neigbours at its midpoint $r=18$\,mm by a flexible, laser-cut elastomeric band of thickness $4$\,mm, whose shape has been optimized such that the band can easily hinge at its anchoring points and such that in-plane compression and stretching of the band dominates its elastic response (Fig.~\ref{fig1}c). The lattice spacing between subsequent rotors is $p=60$\,mm. Assuming that the bending of the band and friction can be neglected, to linear order, the deformations of the elastomeric bands induce a torque on rotor $n$ given by $\tau^{\textrm{elastic}}_n= kr^2\cos^2\theta(  a (a\delta\theta_n-b\delta\theta_{n+1}) -  b (b\delta\theta_n-a\delta\theta_{n-1}))$, where $\delta\theta_n$ is the angular displacement with respect to the initial angle in the reference state, $k=7.8\times 10^{-3}$\,N.m is the spring constant of the rubber band and $a=1.00$ and $b=0.73$ are the geometric parameters defined above.  

In addition, each robotic unit cell is made of a mechanical oscillator, an angular encoder (Broadcom HEDR-55L2-BY09), a DC coreless motor (Motraxx CL1628), a microcontroller (Espressif ESP 32) and a custom-made electronic board that connects these components and ensures power conversion and communication between neighbouring unit cells. Each robotic unit cell uses a local active control loop to apply a strain-dependent torque on rotor $n$ given by $\tau^{\textrm{control loop}}_n= \varepsilon k r^2\cos^2\theta (b (b\delta\theta_n-a\delta\theta_{n-1}) -  a (a\delta\theta_n-b\delta\theta_{n+1}))$, where $\varepsilon$ is a tunable dimensionless non-reciprocal gain parameter. The total torque on the rotors then realizes effective non-reciprocal interactions,
\begin{equation}
\begin{split}
\tau_n=k r^2\cos^2\theta(&-ab(1-\varepsilon)\delta\theta_{n-1}\\&+(a^2(1-\varepsilon)+b^2(1+\varepsilon))\delta\theta_n\\
&-ab(1+\varepsilon)\delta\theta_{n+1}),
\end{split}
\end{equation} 
where $\tau_n=\tau^{\textrm{elastic}}_n+\tau^{\textrm{control loop}}_n$. 
This system precisely realizes the dynamical matrix ${\bm D}$ defined in Eq.~(\ref{D}), and therefore its mechanical response exhibits the bulk-edge correspondence shown by the right zero modes of ${\bm D}$. We record the rotors' instantaneous positions $\delta\theta_n(t)$ via the angular encoders at a resolution $4.4\times 10^{-4}$ rad and sampling frequency $100$ Hz.

\vspace{1eM}
\noindent
\textbf{Measurements of the non-Hermitian winding number}
In order to measure the winding of the bands in the spectrum of the system with periodic boundary conditions, we connect the first and the last rotors with a rigid bar and ball-bearing hinges. To ensure homogeneity of the moments of inertia throughout the system, we add small masses at the end of each rotor.

A rigid pin is attached to each rotor. We impose the initial position of each rotor $\delta\theta_n(t=0)$ away from their equilibrium position as follows: for each wave vector $k$, we manufacture drilled rigid plates with precisely positioned holes, in which the rigid pins can be inserted and that set the initial condition. The initial condition $\delta\theta_n(t=0)=\bar{\delta\theta}\cos k n  $ is chosen such that the overall configuration of the chain is modified from equilibrium according to a specific wave vector ($k=0$ or $k=\pi$). In order to stay in the linear regime as well as within the limit of angular resolution, we impose $\bar{\delta\theta} = 0.21$ rad and $0.04$ rad, respectively, for the measurements on the wave vectors $k=0$ and $k=\pi$. We remove rapidly the drilled plate to let the system freely relax. For each experiment, every unit cell is observed to relax the same way, except for their phase. For each wave vector $k$, we fit the displacement overtime of the far left unit cell to the equation $(A_1 \exp(\lambda_k t)+ A_2 \exp(\lambda'_k t) ) \cos(\omega_k t)$ to deduce the real (imaginary) parts of the eigenfrequencies $\omega_k$ ($\lambda_k$ and $\lambda'_k$), see Figs.~\ref{fig4}bc of the Main Text. From there, we use a discretized version of Eq.~(\ref{eq:NHwinding})
\be
\nu=-\frac{2}{\pi}\left(\arctan\frac{\lambda_\pi-\lambda'_\pi}{2\omega_\pi}-\arctan\frac{\lambda_0-\lambda'_0}{2\omega_0}\right)
\ee
to compute the winding invariant as a function of $\varepsilon$, as shown in Fig.~\ref{fig4}d of the Main Text.

\vspace{1eM}
\noindent
\textbf{Measurements of the edges modes}
We excite the metamaterial at the center rotor ($n=5$) by applying sinusoidal oscillations of amplitude $0.028$ rad and frequency 0.05 Hz over 5 periods of oscillations. We extract the magnitude of the oscillation of each rotor via a Fourier series analysis to produce the data shown in Fig.~\ref{fig4}f-e of the Main Text.

\emph{Acknowledgments.---} {We thank D. Giesen, T. Walstra and T. Weijers for their skilful technical assistance. We are grateful to C. Bender, L. Fu, T. Lubensky, and D. Z. Rocklin, R. Thomale and Z. Wang for insightful discussions and to M.S. Golden for critical review of the manuscript. J.v.W. acknowledge funding from the a Netherlands Organization for Scientific Research (NWO) VIDI grant. C.C. acknowledges funding from the European Research Council grant ERC-StG-Coulais-852587-Extr3Me.}

\bibliographystyle{pnas-new}

\begin{thebibliography}{10}
\expandafter\ifx\csname url\endcsname\relax
  \def\url#1{\texttt{#1}}\fi
\expandafter\ifx\csname urlprefix\endcsname\relax\def\urlprefix{URL }\fi
\providecommand{\bibinfo}[2]{#2}
\providecommand{\eprint}[2][]{\url{#2}}

\bibitem{Nash_PNAS2015}
\bibinfo{author}{Nash, L.~M.} \emph{et~al.}
\newblock \bibinfo{title}{Topological mechanics of gyroscopic metamaterials}.
\newblock \emph{\bibinfo{journal}{Proc. Natl. Ac. Sc. U. S. A.}}
  \textbf{\bibinfo{volume}{112}}, \bibinfo{pages}{14495}
  (\bibinfo{year}{2015}).

\bibitem{Mitchell_NatPhys2018}
\bibinfo{author}{Mitchell, N.~P.}, \bibinfo{author}{Nash, L.~M.},
  \bibinfo{author}{Hexner, D.}, \bibinfo{author}{Turner, A.~M.} \&
  \bibinfo{author}{Irvine, W. T.~M.}
\newblock \bibinfo{title}{Amorphous topological insulators constructed from
  random point sets}.
\newblock \emph{\bibinfo{journal}{Nat. Phys.}} \textbf{\bibinfo{volume}{14}},
  \bibinfo{pages}{380} (\bibinfo{year}{2018}).

\bibitem{Khanikaev_NatComm2015}
\bibinfo{author}{Khanikaev, A.~B.}, \bibinfo{author}{Fleury, R.},
  \bibinfo{author}{Mousavi, S.~H.} \& \bibinfo{author}{Al\`u, A.}
\newblock \bibinfo{title}{Topologically robust sound propagation in an
  angular-momentum-biased graphene-like resonator lattice}.
\newblock \emph{\bibinfo{journal}{Nat. Comm.}} \textbf{\bibinfo{volume}{6}},
  \bibinfo{pages}{8260} (\bibinfo{year}{2015}).

\bibitem{Souslov_PRL2019}
\bibinfo{author}{Souslov, A.}, \bibinfo{author}{Dasbiswas, K.},
  \bibinfo{author}{Fruchart, M.}, \bibinfo{author}{Vaikuntanathan, S.} \&
  \bibinfo{author}{Vitelli, V.}
\newblock \bibinfo{title}{Topological waves in fluids with odd viscosity}.
\newblock \emph{\bibinfo{journal}{Phys. Rev. Lett.}}
  \textbf{\bibinfo{volume}{122}}, \bibinfo{pages}{128001}
  (\bibinfo{year}{2019}).

\bibitem{Delplace_Science2017}
\bibinfo{author}{Delplace, P.}, \bibinfo{author}{Marston, J.~B.} \&
  \bibinfo{author}{Venaille, A.}
\newblock \bibinfo{title}{Topological origin of equatorial waves}.
\newblock \emph{\bibinfo{journal}{Science}} \textbf{\bibinfo{volume}{358}},
  \bibinfo{pages}{1075} (\bibinfo{year}{2017}).

\bibitem{Bandres_Science2018}
\bibinfo{author}{Bandres, M.~A.} \emph{et~al.}
\newblock \bibinfo{title}{Topological insulator laser: Experiments}.
\newblock \emph{\bibinfo{journal}{Science}} \textbf{\bibinfo{volume}{359}},
  \bibinfo{pages}{eaar4005} (\bibinfo{year}{2018}).

\bibitem{Zilberberg_Nature2018}
\bibinfo{author}{Zilberberg, O.} \emph{et~al.}
\newblock \bibinfo{title}{Photonic topological boundary pumping as a probe of
  4d quantum hall physics}.
\newblock \emph{\bibinfo{journal}{Nature}} \textbf{\bibinfo{volume}{553}},
  \bibinfo{pages}{59} (\bibinfo{year}{2018}).

\bibitem{Kraus_PRL2012}
\bibinfo{author}{Kraus, Y.~E.}, \bibinfo{author}{Lahini, Y.},
  \bibinfo{author}{Ringel, Z.}, \bibinfo{author}{Verbin, M.} \&
  \bibinfo{author}{Zilberberg, O.}
\newblock \bibinfo{title}{Topological states and adiabatic pumping in
  quasicrystals}.
\newblock \emph{\bibinfo{journal}{Phys. Rev. Lett.}}
  \textbf{\bibinfo{volume}{109}}, \bibinfo{pages}{106402}
  (\bibinfo{year}{2012}).

\bibitem{Lohse_NatPhys2015}
\bibinfo{author}{Lohse, M.}, \bibinfo{author}{Schweizer, C.},
  \bibinfo{author}{Zilberberg, O.}, \bibinfo{author}{Aidelsburger, M.} \&
  \bibinfo{author}{Bloch, I.}
\newblock \bibinfo{title}{A thouless quantum pump with ultracold bosonic atoms
  in an optical superlattice}.
\newblock \emph{\bibinfo{journal}{Nat. Phys.}} \textbf{\bibinfo{volume}{12}},
  \bibinfo{pages}{350} (\bibinfo{year}{2015}).

\bibitem{Pedro_PRL2019}
\bibinfo{author}{Pedro, R.~P.}, \bibinfo{author}{Paulose, J.},
  \bibinfo{author}{Souslov, A.}, \bibinfo{author}{Dresselhaus, M.} \&
  \bibinfo{author}{Vitelli, V.}
\newblock \bibinfo{title}{Topological protection can arise from thermal
  fluctuations and interactions}.
\newblock \emph{\bibinfo{journal}{Phys. Rev. Lett.}}
  \textbf{\bibinfo{volume}{122}}, \bibinfo{pages}{118001}
  (\bibinfo{year}{2019}).

\bibitem{Zeuner_PRL2015}
\bibinfo{author}{Zeuner, J.~M.} \emph{et~al.}
\newblock \bibinfo{title}{Observation of a topological transition in the bulk
  of a non-hermitian system}.
\newblock \emph{\bibinfo{journal}{Phys. Rev. Lett.}}
  \textbf{\bibinfo{volume}{115}}, \bibinfo{pages}{040402}
  (\bibinfo{year}{2015}).

\bibitem{Gong_PRX2018}
\bibinfo{author}{Gong, Z.} \emph{et~al.}
\newblock \bibinfo{title}{Topological phases of non-hermitian systems}.
\newblock \emph{\bibinfo{journal}{Phys. Rev. X}} \textbf{\bibinfo{volume}{8}},
  \bibinfo{pages}{031079} (\bibinfo{year}{2018}).

\bibitem{Shen_PRL2018}
\bibinfo{author}{Shen, H.}, \bibinfo{author}{Zhen, B.} \& \bibinfo{author}{Fu,
  L.}
\newblock \bibinfo{title}{Topological band theory for non-hermitian
  hamiltonians}.
\newblock \emph{\bibinfo{journal}{Phys. Rev. Lett.}}
  \textbf{\bibinfo{volume}{120}}, \bibinfo{pages}{146402}
  (\bibinfo{year}{2018}).

\bibitem{Ghatak_JPhysCondMat2019}
\bibinfo{author}{Ghatak, A.} \& \bibinfo{author}{Das, T.}
\newblock \bibinfo{title}{New topological invariants in non-hermitian systems}.
\newblock \emph{\bibinfo{journal}{J. Phys.: Cond. Matter}}
  \textbf{\bibinfo{volume}{31}}, \bibinfo{pages}{263001}
  (\bibinfo{year}{2019}).

\bibitem{Yao_PRL2018}
\bibinfo{author}{Yao, S.} \& \bibinfo{author}{Wang, Z.}
\newblock \bibinfo{title}{Edge states and topological invariants of
  non-hermitian systems}.
\newblock \emph{\bibinfo{journal}{Phys. Rev. Lett.}}
  \textbf{\bibinfo{volume}{121}}, \bibinfo{pages}{086803}
  (\bibinfo{year}{2018}).

\bibitem{Kunst_PRL2018}
\bibinfo{author}{Kunst, F.~K.}, \bibinfo{author}{Edvardsson, E.},
  \bibinfo{author}{Budich, J.~C.} \& \bibinfo{author}{Bergholtz, E.~J.}
\newblock \bibinfo{title}{Biorthogonal bulk-boundary correspondence in
  non-hermitian systems}.
\newblock \emph{\bibinfo{journal}{Phys. Rev. Lett.}}
  \textbf{\bibinfo{volume}{121}}, \bibinfo{pages}{026808}
  (\bibinfo{year}{2018}).

\bibitem{Lee_PRB2019}
\bibinfo{author}{Lee, C.~H.} \& \bibinfo{author}{Thomale, R.}
\newblock \bibinfo{title}{Anatomy of skin modes and topology in non-hermitian
  systems}.
\newblock \emph{\bibinfo{journal}{Phys. Rev. B}} \textbf{\bibinfo{volume}{99}},
  \bibinfo{pages}{201103(R)} (\bibinfo{year}{2019}).

\bibitem{Helbig_2019}
\bibinfo{author}{Helbig, T.} \emph{et~al.}
\newblock \bibinfo{title}{Generalized bulk–boundary correspondence in non-Hermitian topolectrical circuits}. \newblock \emph{\bibinfo{journal}{Nat. Phys.}} \textbf{\bibinfo{volume}{16}},
  \bibinfo{pages}{747–750} (\bibinfo{year}{2020}).

\bibitem{Bergholtz_review2019}
\bibinfo{author}{Bergholtz, E.~J.}, \bibinfo{author}{Budich, J.~C.} \&
  \bibinfo{author}{Kunst, F.~K.}
\newblock \bibinfo{title}{Exceptional topology of non-hermitian systems}.
\newblock \emph{\bibinfo{journal}{arXiv:1912.10048}}  (\bibinfo{year}{2019}).

\bibitem{Xiao_NatPhys2020}
\bibinfo{author}{Xiao, L.} \emph{et~al.}
\newblock \bibinfo{title}{Non-hermitian bulk–boundary correspondence in
  quantum dynamics}. \newblock \emph{\bibinfo{journal}{Nat. Phys.}} \textbf{\bibinfo{volume}{16}},
  \bibinfo{pages}{761–766} (\bibinfo{year}{2020}).

\bibitem{MartinezAlvarez_PRB2018}
\bibinfo{author}{Martinez~Alvarez, V.~M.}, \bibinfo{author}{Barrios~Vargas,
  J.~E.} \& \bibinfo{author}{Foa~Torres, L. E.~F.}
\newblock \bibinfo{title}{Non-hermitian robust edge states in one dimension:
  Anomalous localization and eigenspace condensation at exceptional points}.
\newblock \emph{\bibinfo{journal}{Physical Review B}}
  \textbf{\bibinfo{volume}{97}} (\bibinfo{year}{2018}).

\bibitem{Yoshida_arxiv2019}
\bibinfo{author}{Yoshida, T.} \& \bibinfo{author}{Hatsugai, Y.}
\newblock \bibinfo{title}{Exceptional rings protected by emergent symmetry for
  mechanical systems}.
  \newblock \emph{\bibinfo{journal}{Phys. Rev. B}} \textbf{\bibinfo{volume}{100}},
  \bibinfo{pages}{054109} (\bibinfo{year}{2019}).

\bibitem{Hasan_RMP2010}
\bibinfo{author}{Hasan, M.~Z.} \& \bibinfo{author}{Kane, C.~L.}
\newblock \bibinfo{title}{Colloquium: Topological insulators}.
\newblock \emph{\bibinfo{journal}{Rev. Mod. Phys.}}
  \textbf{\bibinfo{volume}{82}}, \bibinfo{pages}{3045} (\bibinfo{year}{2010}).

\bibitem{Fleury_Science2014}
\bibinfo{author}{Fleury, R.}, \bibinfo{author}{Sounas, D.~L.},
  \bibinfo{author}{Sieck, C.~F.}, \bibinfo{author}{Haberman, M.~R.} \&
  \bibinfo{author}{Al\`u, A.}
\newblock \bibinfo{title}{Sound isolation and giant linear nonreciprocity in a
  compact acoustic circulator}.
\newblock \emph{\bibinfo{journal}{Science}} \textbf{\bibinfo{volume}{343}},
  \bibinfo{pages}{516} (\bibinfo{year}{2014}).

\bibitem{Coulais_Nature2017}
\bibinfo{author}{Coulais, C.}, \bibinfo{author}{Sounas, D.} \&
  \bibinfo{author}{Al\`u, A.}
\newblock \bibinfo{title}{Static non-reciprocity in mechanical metamaterials}.
\newblock \emph{\bibinfo{journal}{Nature}} \textbf{\bibinfo{volume}{542}},
  \bibinfo{pages}{461} (\bibinfo{year}{2017}).

\bibitem{Wang_PRL2018}
\bibinfo{author}{Wang, Y.} \emph{et~al.}
\newblock \bibinfo{title}{Observation of nonreciprocal wave propagation in a
  dynamic phononic lattice}.
\newblock \emph{\bibinfo{journal}{Phys. Rev. Lett.}}
  \textbf{\bibinfo{volume}{121}}, \bibinfo{pages}{194301}
  (\bibinfo{year}{2018}).

\bibitem{Kane_NatPhys2014}
\bibinfo{author}{Kane, C.~L.} \& \bibinfo{author}{Lubensky, T.~C.}
\newblock \bibinfo{title}{Topological boundary modes in isostatic lattices}.
\newblock \emph{\bibinfo{journal}{Nat. Phys.}} \textbf{\bibinfo{volume}{10}},
  \bibinfo{pages}{39} (\bibinfo{year}{2014}).
  
\bibitem{Brandenbourger_arxiv2019}
\bibinfo{author}{Brandenbourger, M.}, \bibinfo{author}{Locsin, X.},
  \bibinfo{author}{Lerner, E.} \& \bibinfo{author}{Coulais, C.}
\newblock \bibinfo{title}{Non-reciprocal robotic metamaterials}.
\newblock \emph{\bibinfo{journal}{Nat. Commun.}} \textbf{\bibinfo{volume}{10}},
  \bibinfo{pages}{4608} (\bibinfo{year}{2019}).


\bibitem{Lubensky_RepProgPhys2015}
\bibinfo{author}{Lubensky, T.~C.}, \bibinfo{author}{Kane, C.~L.},
  \bibinfo{author}{Mao, X.}, \bibinfo{author}{Souslov, A.} \&
  \bibinfo{author}{Sun, K.}
\newblock \bibinfo{title}{Phonons and elasticity in critically coordinated
  lattices}.
\newblock \emph{\bibinfo{journal}{Rep. Prog. Phys.}}
  \textbf{\bibinfo{volume}{78}}, \bibinfo{pages}{073901}
  (\bibinfo{year}{2015}).

\bibitem{Huber_NatPhys2016}
\bibinfo{author}{Huber, S.~D.}
\newblock \bibinfo{title}{Topological mechanics}.
\newblock \emph{\bibinfo{journal}{Nat. Phys.}} \textbf{\bibinfo{volume}{12}},
  \bibinfo{pages}{621} (\bibinfo{year}{2016}).

\bibitem{Susstrunk_PNAS2016}
\bibinfo{author}{Susstrunk, R.} \& \bibinfo{author}{Huber, S.~D.}
\newblock \bibinfo{title}{Classification of topological phonons in linear
  mechanical metamaterials}.
\newblock \emph{\bibinfo{journal}{Proc. Natl. Ac. Sc. U. S. A.}}
  \textbf{\bibinfo{volume}{113}}, \bibinfo{pages}{E4767}
  (\bibinfo{year}{2016}).
  
\bibitem{history_topo}
    \bibinfo{author}{Zak, J.}
\newblock \bibinfo{title}{Symmetry criterion for surface states in solids}.
\newblock \emph{\bibinfo{journal}{Phys. Rev. B}} \textbf{\bibinfo{volume}{32}},
  \bibinfo{pages}{2218} (\bibinfo{year}{1985}).
  
\bibitem{McDonald_PRX2018}
\bibinfo{author}{McDonald, A.}, \bibinfo{author}{Pereg-Barnea, T.} \&
  \bibinfo{author}{Clerk, A.~A.}
\newblock \bibinfo{title}{Phase-dependent chiral transport and effective
  non-hermitian dynamics in a bosonic kitaev-majorana chain}.
\newblock \emph{\bibinfo{journal}{Phys. Rev. X}} \textbf{\bibinfo{volume}{8}},
  \bibinfo{pages}{041031} (\bibinfo{year}{2018}).

\bibitem{Mostafazadeh_IJGMMP2010}
\bibinfo{author}{Mostafazadeh, A.}
\newblock \bibinfo{title}{Pseudo-hermitian representation of quantum
  mechanics}.
\newblock \emph{\bibinfo{journal}{Int. J. Geom. Meth. Mod. Phys.}}
  \textbf{\bibinfo{volume}{07}}, \bibinfo{pages}{1191} (\bibinfo{year}{2010}).

\bibitem{Borgnia_arxiv2019}
\bibinfo{author}{Borgnia, D.~S.}, \bibinfo{author}{Kruchkov, A.~J.} \&
  \bibinfo{author}{Slager, R.-J.}
\newblock \bibinfo{title}{Non-Hermitian Boundary Modes and Topology}.
\newblock \emph{\bibinfo{journal}{Phys. Rev. Lett.}}   \textbf{\bibinfo{volume}{124}}  \bibinfo{pages}{056802}  (\bibinfo{year}{2020}).

\bibitem{Schomerus}
\bibinfo{author}{Schomerus, H.}
\newblock \bibinfo{title}{Nonreciprocal response theory of non-hermitian
  mechanical metamaterials: Response phase transition from the skin effect of
  zero modes}.
\newblock \emph{\bibinfo{journal}{Phys. Rev. Res.}}
  \textbf{\bibinfo{volume}{2}}  \bibinfo{pages}{013058} (\bibinfo{year}{2020}).

\end{thebibliography}

\end{document}